\documentclass[%
 amsmath,amssymb,
 draft,
 grl,
]{agutex}

\usepackage[dvips]{graphicx}
\usepackage{dcolumn}
\usepackage{bm}

\setkeys{Gin}{draft=false}

\usepackage{color}
\usepackage{amstext}
\usepackage{amsmath}

\authoraddr{Corresponding author: Colby Haggerty, 
Bartol Research Institute, Department of Physics and
Astronomy, University of Delaware, Newark, Delaware, USA
(colbych@udel.edu)}

\lefthead{Haggerty et al.}
\righthead{Partition of heating in reconnection}

\begin{document}

\title{The competition of electron and ion heating during magnetic reconnection} 

\authors{C. C. Haggerty,\altaffilmark{1}
M. A. Shay,\altaffilmark{1}
J. F. Drake,\altaffilmark{2}
T. D. Phan,\altaffilmark{3}
C. T. McHugh\altaffilmark{1}}

\altaffiltext{1}{Bartol Research Institute, Department of Physics and
  Astronomy, University of Delaware, Newark, DE 19716, USA}

\altaffiltext{2}{Department of Physics and the Institute for Physical Science and Technology, University of Maryland, College Park, Maryland 20742, USA}

\altaffiltext{3}{Space Sciences Laboratory, University of California, Berkeley,
California 94720, USA}



\begin{abstract}

The physical processes that control the partition of released magnetic
energy between electrons and ions during reconnection is explored
through particle-in-cell simulations and analytical techniques. We
demonstrate that the development of a large-scale parallel electric
field and its associated potential controls the relative heating of
electrons and ions. The potential develops to restrain
heated exhaust electrons and enhances their heating by confining electrons in the region where magnetic energy is released. Simultaneously
the potential slows ions entering the exhaust below the
Alfv\'enic speed expected from the traditional counterstreaming
picture of ion heating. Unexpectedly, the magnitude of the potential and therefore the relative partition of energy between
electrons and ions is not a constant but rather depends on
the upstream parameters and specifically the upstream electron
normalized temperature (electron beta).  These findings  suggest that the fraction of magnetic energy converted into the total thermal energy may be independent of upstream parameters.

\end{abstract}



\begin{article}

\section{Introduction}
Magnetic reconnection is a universal plasma
process which converts stored magnetic energy into particle
energy. The process is believed to be important in many astrophysical,
solar, geophysical, and laboratory contexts.  A principle topic in
reconnection physics is the mechanism by which magnetic energy is
partitioned into electron and ion thermal energy. A measure of this
partition is the relative fraction of the available magnetic energy
per particle $W = B_{rup}^2/(4\pi\, m_i\, n_{up})=m_i\, c_{Aup}^2$ (or its
asymmetric generalization $m_i\,c_{A,asym}^2 $  \citep{Phan13,Shay14}) that goes to each class of
particle; the supscript "up" denotes the upstream value and 
$B_{rup}$ is the reconnecting component of the magnetic field.

Ion thermal energy often makes up a large fraction of the released
magnetic energy during magnetic reconnection in both in the
magnetosphere \citep{Eastwood13,Phan14} and the laboratory
\citep{Yamada14}. In the reconnection exhaust, where most magnetic
energy is released, ion heating takes the form of interpenetrating
beams
\citep{Cowley82,Krauss-Varban95,Nakabayashi97,Hoshino98,Gosling05,Lottermoser98,Stark05,Wygant05,Phan07},
which are generated through Fermi reflection in the outflowing,
contracting magnetic fields. The predicted counterstreaming velocity
is twice the exhaust velocity $c_{Aup}$ in the case of antiparallel
reconnection even in the presence of Hall magnetic and electric fields
\citep{Drake09}. The expected ion temperature increase based on such a
simple picture is $\Delta T_i = 0.33\ m_ic_{Aup}^2 = 0.33\,W$. However, in solar
wind and magnetopause observations the ion temperature increments are
significantly lower than expected, $\Delta T_i\sim 0.13\,W,$ but exhibit the expected scaling with parameters \citep{Drake09,Phan14}.

The scaling of electron heating is much more challenging to understand
because the single-pass Fermi reflection yields only a small increase
in the electron temperature. Nevertheless, magnetopause
observations for electrons yield a similar scaling $\Delta T_e\sim
0.017\,W$ although with significantly less heating compared to the
ions \citep{Phan13}. Simulations also yield this scaling \citep{Shay14} although the
electron heating mechanism remains under debate
\citep{Haggerty14,Egedal14}. 

Thus, it is important not only to establish the explicit mechanisms
for electron and ion heating during reconnection but also to determine
whether the partition of energy between the two species is a universal
relation or varies with parameters. We demonstrate here through a set
of comprehensive computer simulations and analytic methods that the
large-scale parallel potential that develops within the reconnection
exhaust controls and links together both electron and ion heating and
regulates the partition of released magnetic energy. The development
of this potential within the exhaust to prevent the escape of hot
electrons has been well-established \citep{Egedal08} and enables
electrons to undergo repeated Fermi reflections within the
reconnection exhaust. In the present paper we identify the mechanism
that ultimately limits electron energy gain. The spatial variation of the potential propagates
outward from the exhaust as a component of a slow shock
\citep{Liu12}. The electron temperature and the associated shock
velocity increase until the velocity matches that of the Alfv\'enic
exhaust. At this point electron energy gain through Fermi reflection
ends since the bounce length of electrons trapped in the exhaust no
longer decreases with time as they propagate downstream. At the same
time that the potential serves to facilitate electron energy gain, it
suppresses ion heating: the parallel streaming velocity of ions
injected into the exhaust from upstream is reduced below the Alfv\'en
speed by the potential so that the counterstreaming velocity of ions
is less than $2c_{Aup}$. Thus, the strength of the potential regulates
the relative heating of electrons and ions. We show that the potential
increases with increasing upstream electron temperature $T_{eup}$ and
that $\Delta T_e$ can actually exceed $\Delta T_i$ -- the partition of
electron and ion heating measured in the magnetosphere
\citep{Eastwood13,Phan13,Phan14} and laboratory experiments
\citep{Yamada14} is not universal.
However, the total heating is unaffected by the potential and the fraction of magnetic energy converted into thermal energy is constant for the simulations performed, with $\Delta (T_i + T_e) \approx 0.15\ m_i\,c_{Aup}^2 = 0.15\,W.$ 
Remarkably, despite the numerous differences between the simulations and observations, this slope is the
same as the \citet{Phan13,Phan14} measurement of total heating at the Earth's magnetopause.


\section{Simulations:} We use the PIC code P3D\citep{Zeiler02} to
perform simulations in 2.5 dimensions of collisionless antiparallel
(no guide field) reconnection. Magnetic field strengths and particle
number densities are normalized to \(B_{0}\) and \(n_{0}\),
respectively.  Lengths are normalized to the ion inertial length
\(d_{i0}=c/\omega_{pi0}\) at the reference density \(n_{0}\), time to
the ion cyclotron time \(\Omega_{ci0}^{-1}=(eB_{0}/m_{i}c)^{-1},\) and
velocities to the Alfv\'en speed
\(c_{A0}=\sqrt{B_{0}^{2}/(4\pi\,m_{i}\,n_{0})}\). Electric fields and
temperatures are normalized to \(E_{0}=c_{A0}B_{0}/c\) and
\(T_{0}=m_{i}c_{A0}^{2}\), respectively. In the simulation coordinate
system the reconnection outflows are along \(\hat{x}\) and the inflows
are along \(\hat{y}\).  Simulations are performed in a periodic domain
with a system size of \(L_{x} \times L_y = 204.8\, d_{i0} \times
102.4\, d_{i0} \), and 100 particles per grid in the inflow
region. Simulation parameters, which are given in the table in the Supplementary Material, included ion-to-electron mass-ratios of $25$ and $100$ and a variety of upstream initial temperatures and magnetic fields.
The initial conditions are a double current sheet\citep{Shay07}.

A small magnetic perturbation is used to initiate reconnection.  Each
simulation is evolved until reconnection reaches a steady state, and
then for analysis purposes during this steady period the simulation
data is time averaged over 100 particle time steps, which is typically
on the order of 50 electron plasma wave periods \(\omega_{pe}^{-1}\).

\section{Overview of electron and ion heating:} We first present an overview of electron and ion heating as measured
in the simulations. The temperature of electrons and ions each
increase with the distance downstream of the x-line in the exhaust
until it approaches a constant. This behavior has already been
discussed in detail for electrons \citep{Shay14} and is discussed more
fully in the Supplementary Material for the ions.
To determine $\Delta T_i$ and $\Delta T_e$ in a given simulation we
average $T_i$ and $T_e$ over a region downstream and then subtract the
inflow temperature. Details of how this average is computed for ions
are found in the Supplementary Material. In Fig.~\ref{Fig:overview} we
present an overview of (a) electron, (b) ion and (c) the total
temperature increments versus $m_ic_{Aup}^2$. The red triangles
correspond to high upstream electron temperature $T_e/T_i=9$. As expected, the sum of the electron and ion heating increments scale with the available magnetic
energy per particle with an approximate slope of 
\(\Delta (T_i + T_e) \approx 0.15\ m_i\,c_{Aup}^2 = 0.15\,W.\) This slope is the
same as measured in observations of electron and ion heating in reconnection exhausts at the
Earth's magnetopause \citep{Phan13,Phan14}. Surprisingly, however, the individual electron and ion
temperature increments in Fig.~\ref{Fig:overview} have a larger spread related to the upstream electron temperature. The electron heating is generally
significantly below that of the ions, as in the observational data
\citep{Phan13}. The exceptions are the runs with high electron
temperature upstream, which produce enhanced electron heating and reduced
ion heating with the electron heating significantly greater than the ion
heating. These simulations therefore demonstrate the
parameter-dependence of energy partition between electrons and
ions. In the remainder of the manuscript we explore the mechanisms
that control the heating of both species, starting with the ions.

\begin{figure}
\includegraphics[width=3.5in]{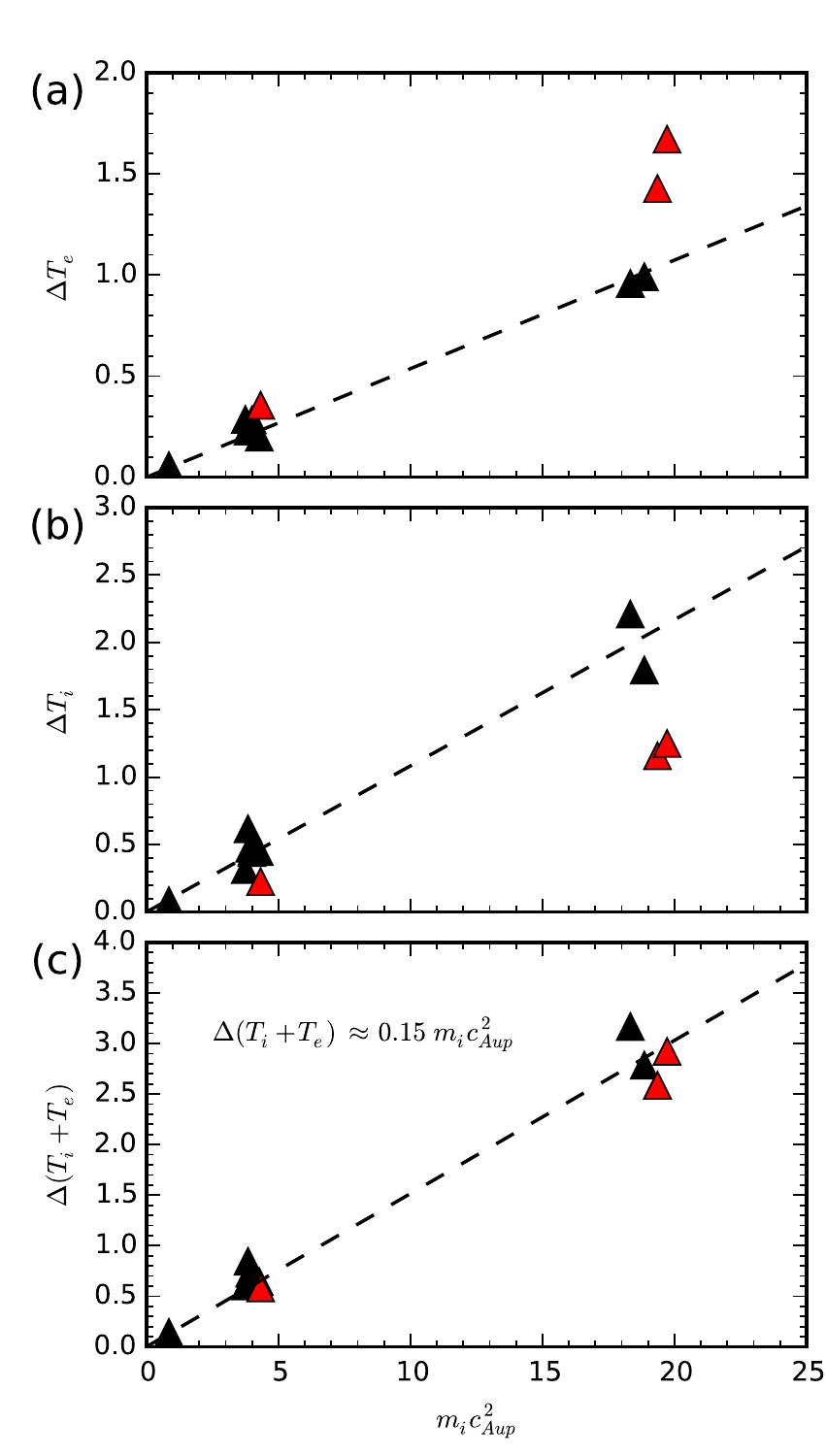}
\caption{Overview of electron and ion heating: (a) $\Delta T_e$, (b) $\Delta T_i$ and (c) $\Delta (T_e+T_i)$ versus $m_ic_{Aup}^2$. The three red triangles have upstream $T_e/T_i = 9$. The change in total temperature appears insensitive to  $T_{eup}$ with $\Delta (T_e+T_i) \approx 0.15\ m_{i}c_{Aup}^2$
}\label{Fig:overview}
\end{figure}


\begin{figure}[!t]
\begin{center}
\includegraphics[width=3.5in]{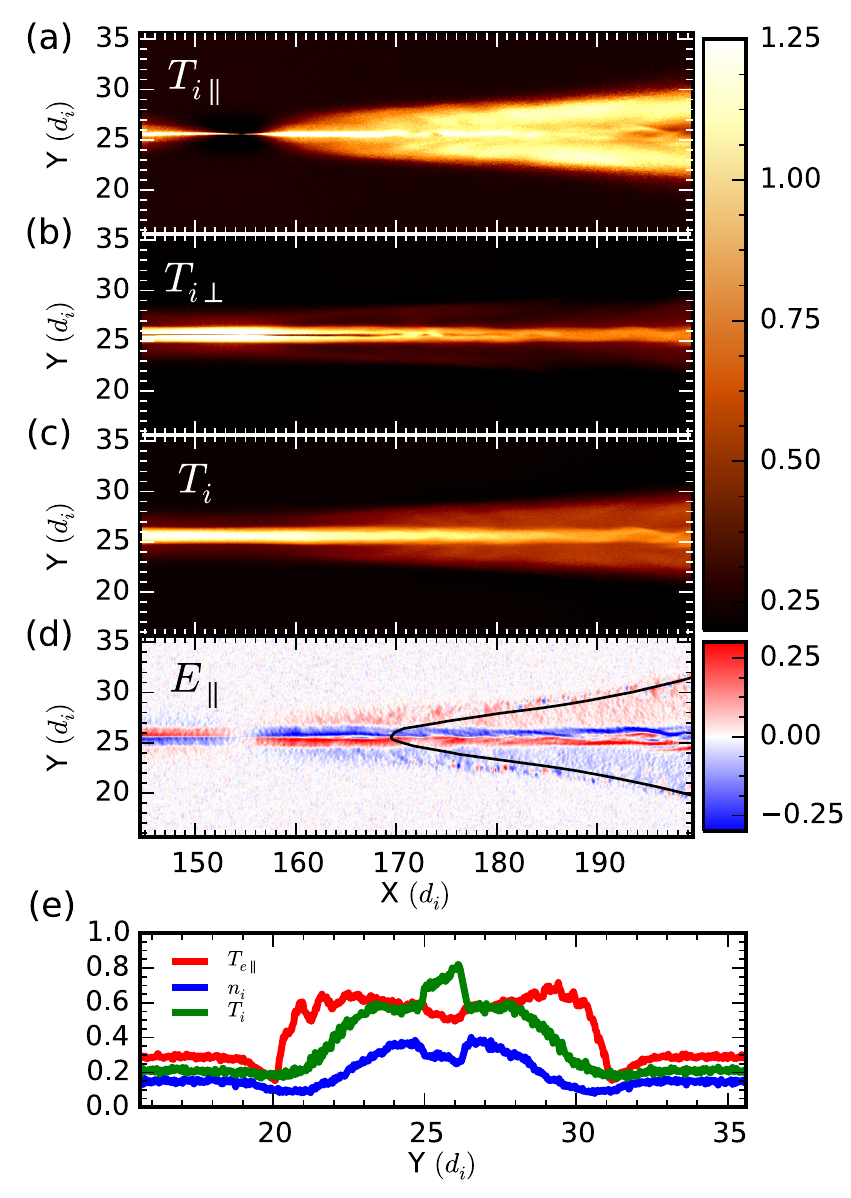}
\end{center}
\caption{Exhaust structure for a typical simulation with \(B = 1.0,\ n = 0.2,\ T_i = T_e = 0.25\), and $m_i/m_e = 100$ (a) $T_{i\parallel}$, (b) $T_{i\perp}$ and
  (c) $T_{i}$ using the same color scale; (d) Spatially smoothed
  $E_\parallel$ with single field line in black. Note that the large scale parallel electric
  field that fills the exhaust, as opposed to the small scale electric field at the midplane,
plays a key role in modifying the heating. (e) $n$, $T_i,$ and
  $T_{e\parallel}$ along $Y$ at $X=194.5$.}
\label{Fig:Temperature-determination}
\end{figure}

Shown in Figs.~\ref{Fig:Temperature-determination}a-c are the ion
parallel temperature \(T_{i\parallel}\), perpendicular temperature
\(T_{i\perp}\) and total temperature \((T_i \equiv \left[ T_{i\parallel} + 2 T_{i\perp} \right] /3) \). Downstream of the x-line \(T_{i\parallel}\)
increases and broadens in the inflow direction to fill the
exhaust. The band of \(T_{i\perp}\) at the midplane of the exhaust is
produced by the Speiser orbits of the ions \citep{Speiser65,Drake09}. As with the
electrons, the total ion temperature asymptotes to a constant
downstream \citep{Shay14}.  The underlying mechanism for ion
heating well downstream of the x-line was outlined by \cite{Drake09}. In this downstream region,
${\bf E}_\perp = 0$ in the reference frame moving with the reconnected magnetic field lines.
This includes both the reconnecting and Hall electric fields as shown in Fig.~3 of \cite{Drake09}.
Within the ion diffusion region, however, the strong normal electric field cannot be transformed
away~\citep{Wygant05}. In
this moving frame the cold ion population enters the reconnection
exhaust with a parallel velocity equal to the field line velocity
$v_0$. The ions reach the midplane, undergo an energy-conserving
reflection, and then travel back out along the field line. The
reflected population mixes with cold incoming ions creating
counter-streaming beams and a temperature increment of $\Delta T_i \approx
\Delta T_{i\parallel}/3 \approx m_{i} v_0^2/3.$ In order to test
this prediction we directly measure the field line velocity $v_0
\approx -c\,E_z/B_y$, which asymptotes to the ion outflow velocity
$v_{ix}$ in the downstream region.
\begin{figure}
\begin{center}
\includegraphics[width=3.5in]{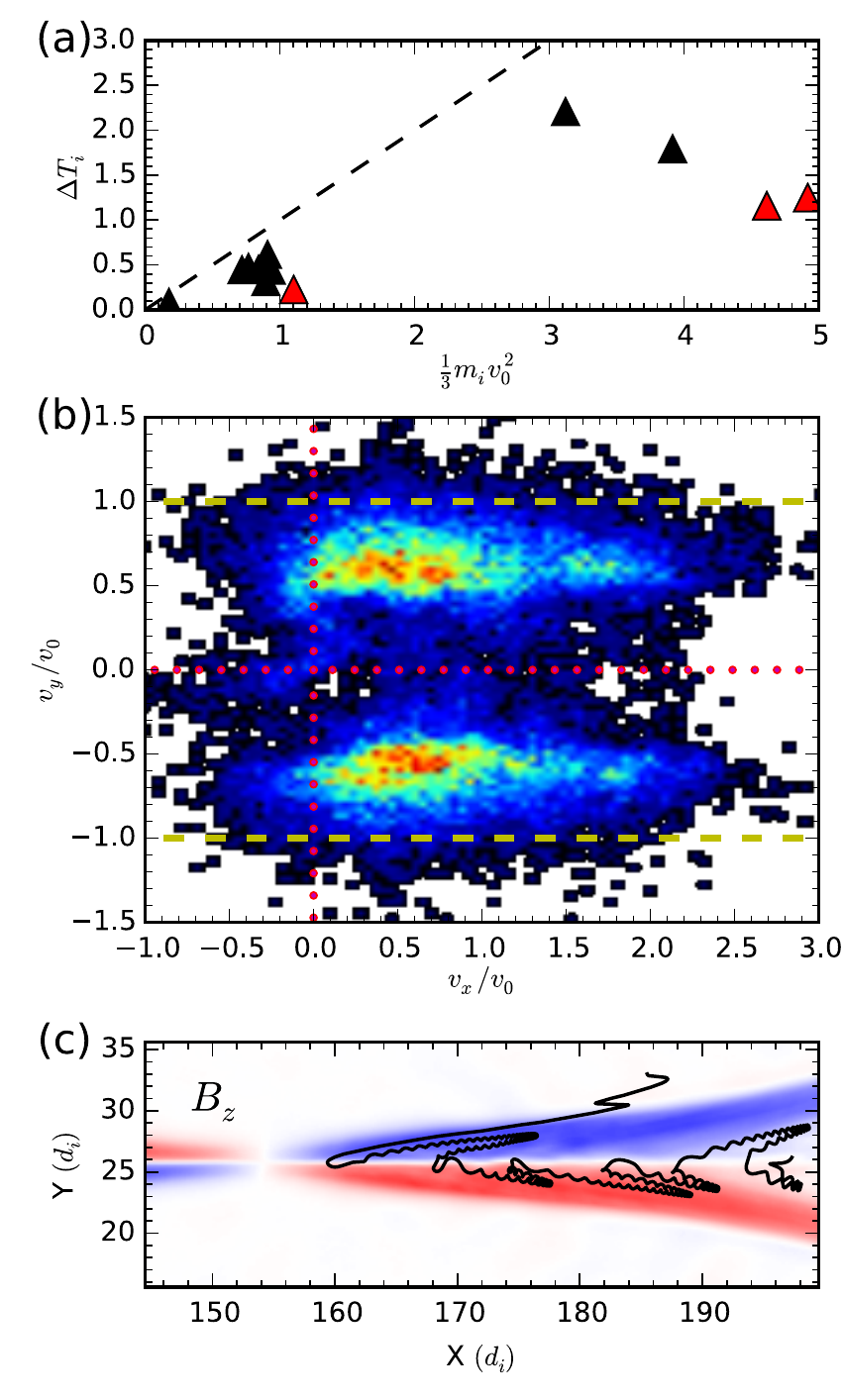}
\end{center}
\caption{Test of the basic counterstreaming ion model. (a) \(\Delta T_i\) versus the theoretical prediction
$m_i\,v_0^2/3$ (the dashed line has a slope of 1). The three red triangles
have upstream \(T_e/T_{i} = 9.\) (b) Ion distribution function around \((X,Y)=(25.6,190)\) from the simulation in Fig.~\ref{Fig:Temperature-determination}.  Velocities are normalized to \(v_0\) (the asymptotic
field line velocity), with doted red lines showing \(v_{x,y} = 0\) and dashed yellow lines showing \(v_y = v_0\). (c) 2D trajectory of a test particle (electron)
entering the reconnection exhaust plotted over Bz. The particle was initialized upstream with the local ExB velocity, and shows the typical trajectory of an electron in the reconnection exhaust. The particle is evolved in the fields from the time averaged simulation using the Boris algorithm.}
\label{Fig:Ti-too-small}
\end{figure}
The prediction of $\Delta T_i = m_i\,v_0^2/3$ is
tested in Fig.~\ref{Fig:Ti-too-small}a. The points roughly scale with
\(m_i\, v_0^2/3\), but there are two significant differences relative
to the theoretical value: (1) There are outlier points leading to a
large spread of the data, and (2) all of the data points are
substantially below the theoretical prediction (line of slope = 1 as
indicated by the dashed black line). In Fig.~\ref{Fig:Ti-too-small}b,
examination of the ion distribution function integrated along \(v_z\)
around \((X,Y) = (190,25.6)\) reveals that the beaming velocities are
significantly less than \(v_0\). The magnetic field points along
\(\hat{y},\) and two field-aligned counterstreaming populations
straddle \(v_y = 0 \) but well within the region \(|v_y/v_0| < 1
\). 

We now show that the reduction in ion heating is a consequence of the
large-scale potential that confines the hot electrons (see cuts of
$T_{e\parallel}$ and $n$ in
Fig.~\ref{Fig:Temperature-determination}e) in the exhaust. In order to
maintain electron force balance along the magnetic field, a
large-scale, although relativity small magnitude, parallel electric
field arises (Fig.~\ref{Fig:Temperature-determination}d). The
$E_\parallel$ fills the exhaust and points away from the
midplane. This electric field and associated potential slows down
inflowing ions leading to a reduced ion beam velocity and a reduced
\(\Delta T_i\). 

Note that in Fig.~\ref{Fig:Temperature-determination}d there is an inverted $E_\parallel$ structure
straddling the midplane that is not to be confused with the larger scale parallel field discussed
above. This smaller scale parallel electric field is connected with the outer electron diffusion region associated with the super-Alfv\'enic electron jet~\citep{Shay07},
and does not couple to the ions, which are unmagnetized at these small scales. For that reason,
the effect of this electron scale parallel electric field is not included in our analysis.

\begin{figure}
\begin{center}
\includegraphics[width=3.5in]{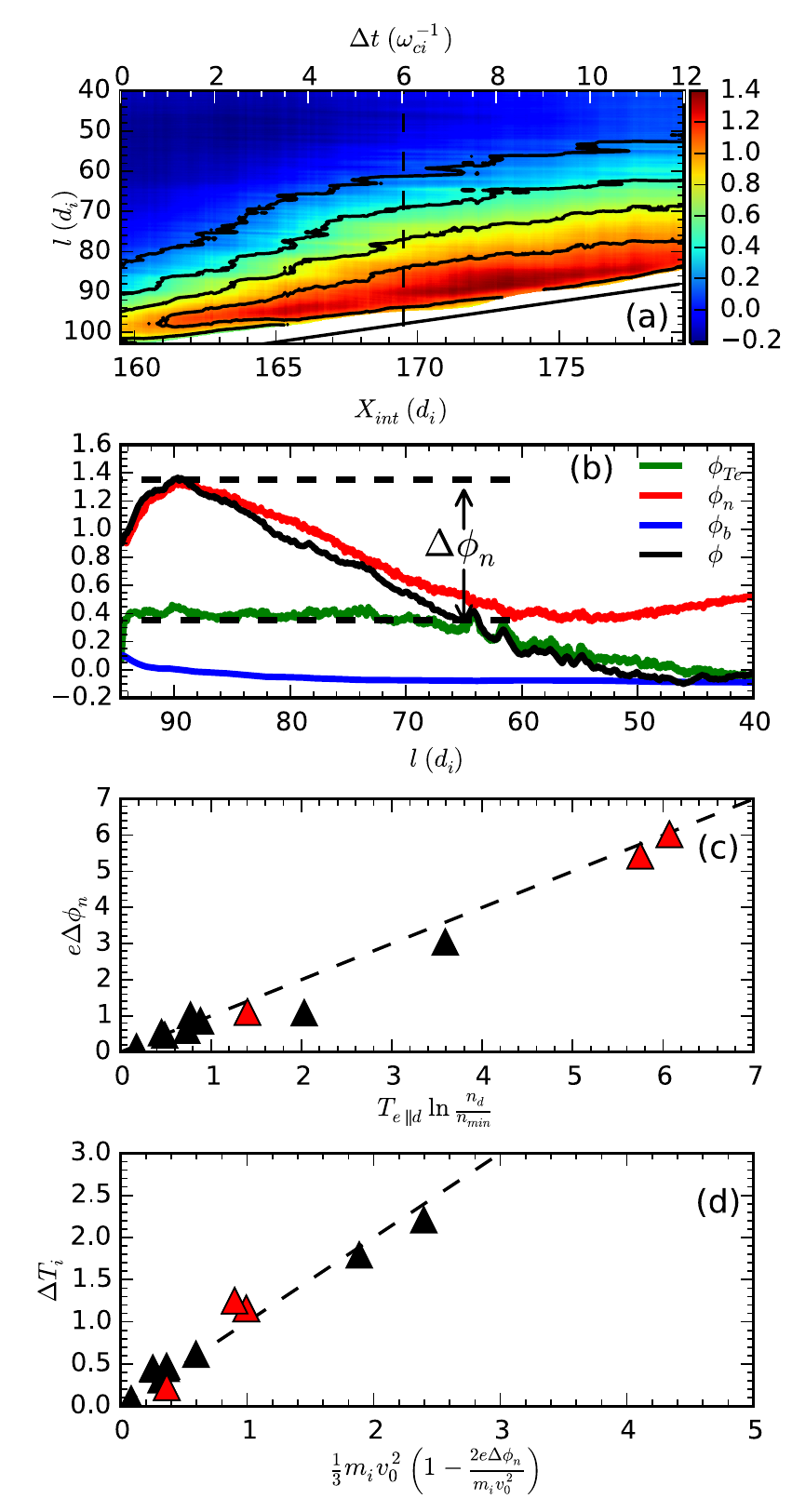}
\end{center}
\caption{(a) The shortening of field lines in the expanding exhaust
  and the field-aligned propagation of the spatial variation of  $\phi$ for the same simulation as Fig~\ref{Fig:Temperature-determination}. Shown is $\phi$ as a
  function of distance $l$ along a field line (with $l=0$ at the \(X\) value of the middle of the island \((X = 256.9)\)) and $X_{int}$ the intercept
  location of the field line at the midplane of the exhaust. The time
  axis is defined by $\Delta t=\Delta X_{int}/v_0$ where $v_0$ is  the asymptotic field line velocity. $\phi$ is taken to be zero at $l=0$. (b)
  $\phi$, $\phi_{Te}$, $\phi_{n}$, and $\phi_{B}$ versus $l$ along the solid
  black magnetic field line in
  Fig.~\ref{Fig:Temperature-determination}d with \(l=0\) defined as in (a). (c)
  $e\Delta\phi_n$ versus $T_{e\parallel d}ln(n_d/n_{min})$. (d)\(\Delta T_i\) versus the predicted temperature increment including
  the effect of $\Delta \phi_n$ (Eq.~\ref{eqn:Ti-reduction}).  
\label{Fig:phi-effect-on-Ti}
}
\end{figure}

To calculate the impact of the large scale parallel electric field and the associated potential on the ions, it is necessary
to  understand both its amplitude and  space-time
structure.  The spatial variation of the potential propagates as a component of the exhaust boundary moving outward away from the midplane. This exhaust boundary takes the form of super-slow to sub-slow transition rather than a switch-off shock because of the strong temperature anisotropy that develops in collisionless reconnection\citep{Liu12}. We determine this velocity directly from
the simulation by calculating the
potential $\phi$ by integrating $E_\parallel$ along the magnetic field. We take $\phi =0$ at the $X$ value of the middle of the island \((X = 256.4)\), where the
distance along the field line $l$ is also taken to be zero. In
Fig.~\ref{Fig:phi-effect-on-Ti}a we plot $\phi$ versus $l$ and the
$X$-intercept of the field line with the midplane of the exhaust, denoted as \(X_{int}\). Only
the portion above the exhaust midplane is shown so that the expansion
of the white zone with distance downstream measures the rate of
shortening of the field line (using the time axis which is defined by
$\Delta t=\Delta X/v_0$). The boundary of the white zone parallels the solid line in
the white zone, which marks the exhaust velocity $v_0$, so field line
shortening is at the velocity $v_0$ as expected. The more important
result of Fig.~\ref{Fig:phi-effect-on-Ti} is that the contours of
$\phi$ parallel the boundary of the white zone which means that the
expansion velocity of the potential is $v_0$, the same as the
shortening rate of the field lines. This is a crucial result that will
enable us to explicitly calculate ion heating and impose limits on
electron heating.

We analytically calculate the magnitude of $\phi$ from the 
parallel electric field, which follows from electron force
balance:\
\begin{equation} 
eE_\parallel = -\nabla_b\, T_{e\parallel} - T_{e\parallel}\, \nabla_b \ln{n} +
 (\,T_{e\parallel} - T_{e\perp}\,)\nabla_b \ln{B}, 
 \label{eqn:e-pressure-balance}
\end{equation}
where \( \nabla_b = (\mathbf{B}/B) \cdot \boldsymbol{\nabla}\).  The
potential \(\phi, \) is then given by \(\phi \equiv - \int
{E_\parallel}\,dl,\). Integrating Eq.~\ref{eqn:e-pressure-balance} and multiply both side by \(-1\) yields $\phi = \phi_{Te} + \phi_{n} +
\phi_{B},$ where the subscript represents the quantity acted on by the
gradient, i.e., 
\begin{equation}
\phi_n \equiv \int (T_{e\parallel}/e)\,(\nabla_b\,\ln n )\,dl.
\label{eqn:phi_n} 
\end{equation}

In Fig.~\ref{Fig:phi-effect-on-Ti}b, these potentials
are plotted along the solid black field line shown in
Fig.~\ref{Fig:Temperature-determination}d. $\phi_{Te}, \phi_{n},$ and
$\phi_{B}$ have different constants added to aid in their comparison
with $\phi$.
$\phi$ increases from the inflow region to the exhaust, reaching its
maximum value just outside the midplane. We have found through test
particle simulations that as the ions enter the
exhaust only $\phi_{n}$ significantly modifies the
ion beam velocity (and therefore $\Delta T_i$). $\phi_B$ is small. $\phi_{Te}$ is significant in a narrow
region at the edge of the exhaust but because it is so localized and
because there is a large transverse electric field in this region, 
test particle trajectories provided in the  supplementary material reveal  that the ions cross this region
transverse to ${\bf B}$ and don't respond to $\phi_{Te}$. The dip in
$\phi$ at the midplane of the exhaust is similarly unimportant since
it only affects the ion temperature within a narrow region that
occupies a decreasingly small fraction of the exhaust with distance
downstream.

Thus, $\phi_n$ has the greatest impact on $\Delta T_i$. To calculate
$\Delta T_i$ we therefore need to evaluate the jump in $\phi_n$ across
the exhaust. Since $T_{e\parallel}$ is nearly constant across the exhaust (Fig.~\ref{Fig:Temperature-determination}e) we can replace it by its average value $T_{e\parallel d}$ in the integral in Eq.~\ref{eqn:phi_n}. The density varies from a minimum $n_{min}$ at the
exhaust boundary to a maximum $n_d$ in the middle of the exhaust so the jump in $\phi_n$ across the exhaust $\Delta\phi_n$ is given by
\begin{equation}
e\Delta\phi_n\approx T_{e\parallel d}\,\ln(n_d/n_{min}).
\label{eqn:delta-phi_n}
\end{equation}

The jump $\Delta\phi_n$ is marked in the simulation data in
Fig.~\ref{Fig:phi-effect-on-Ti}b. In Fig.~\ref{Fig:phi-effect-on-Ti}c
we plot the value of $\Delta\phi_n$ measured from the simulation against the values
 from Eq.~\ref{eqn:delta-phi_n}. The agreement is excellent. Note the
large value of the potential for the simulations with high value of
upstream $T_e/T_i$ (red triangles). We can now extend the model of
\citet{Drake09} to include the effect of \(\phi_n \)
to obtain a more accurate ion heating prediction. In a frame moving
with the field line the potential is also unchanging since its outflow
velocity is also $v_0$. In this frame the incoming population will be
slowed down from the field line velocity (\(v_0\)) to the exhaust beam
velocity (\(v_d\)). These slower ions mix with incoming ions from the
other side of the midplane, leading to counterstreaming beams and a
temperature increment of \(\Delta T_i = m_i v_d^2/3\). In the frame of
the potential the ion energy is conserved so we can calculate $v_d$
directly from
$\frac{1}{2}\,m_i\,v_0^2 - e\Delta\phi_n = \frac{1}{2}\,m_i\,v_d^2.$
Solving for \(v_d\) and substituting in \(\Delta T_i,\) we find 
\begin{equation}
\Delta T_i = \frac{m_i\,v_0^2}{3} \left(1 - \frac{2\,e\, \Delta \phi_n}{m_i\,v_0^2}
\right)
\label{eqn:Ti-reduction}
\end{equation}


In Fig.~\ref{Fig:phi-effect-on-Ti}d we insert the measured $\Delta \phi_n$  from the simulation in
Eq.~\ref{eqn:Ti-reduction} and compare the prediction with the
measured ion heating in the simulations. The spread in the data is
markedly reduced compared with that in Fig.~\ref{Fig:Ti-too-small}a:
all of the points now straddle a line with a slope of 1.  Most
revealing is the change in position of the \(T_e/T_i = 9\) simulations
which are denoted by  red triangles in
Figs.~\ref{Fig:Ti-too-small} and \ref{Fig:phi-effect-on-Ti}. These
simulations have large \(\Delta \phi_n\) which significantly reduces
the ion beam velocity and the corresponding ion temperature increment. Thus,
electrons, through the self-generated potential have a strong impact
on ion heating.

A question remains as to why previous observational studies measure an
ion increment $\Delta T_i \propto m_i\,v_{0}^2$~\citep{Drake09,Phan14},
even though such a scaling is not implied by
Eq.~\ref{eqn:Ti-reduction} due to the presence of the potential $\Delta
\phi_n$. $\Delta \phi_n$ depends on both $\Delta
T_{e\parallel},$ and $T_{eup}$ in
Eq.~\ref{eqn:phi_n}; the logarithm of the density
compression ratio is not expected to vary significantly with upstream
conditions for symmetric reconnection. For a significant variation of upstream properties, $\Delta T_{e\parallel}$ has been shown to scale
with $m_i\, v_0^2 \approx m_ic_{Aup}^2$~\citep{Phan13,Shay14}. Any
deviation from the $m_i\, v_0^2$ scaling, therefore, is linked to $T_{eup}/(m_i
c_{Aup}^2) \approx \beta_{eup}/2$. As long as the electron heating is sufficiently
strong compared to the upstream temperature, $\Delta T_e$ should dominate, and we recover both the
observational scaling and the scaling of the black triangles in
Fig.~\ref{Fig:Ti-too-small}a.

Since weak parallel electric fields are impossible to directly measure
with in situ satellite measurements, the analytic expression for
$e\Delta\phi_n$ in Eq.~\ref{eqn:delta-phi_n} can be used to evaluate $\Delta
T_i$ in Eq.~\ref{eqn:Ti-reduction} to compare with observations. In addition, the prediction can be further simplified by using the approximation $v_0 \approx c_{Aup}$.


We now discuss the impact of the potential on electron heating. It has been
shown that the dominant driver of electron heating
during anti-parallel reconnection is Fermi reflection
\citep{Dahlin14}. In the absence of scattering, electron energy gain is
mostly along the local magnetic field. On the other hand, a single
Fermi reflection of electrons in the reconnection exhaust is not
sufficient to drive significant electron energy gain. Electrons can
gain energy through multiple Fermi reflections during multi x-line
reconnection \citep{Drake06,Oka10,Drake13} or in a single x-line reconnection as a result of the
potential $\phi$, which acts to confine electrons within the
reconnection exhaust \citep{Egedal08}. What limits
the electron temperature within the exhaust $T_{ed\parallel}$ and
therefore the potential (Eq.~\ref{eqn:phi_n}) has not been
established. Electrons can continue to gain energy in a single exhaust
by repeatedly reflecting off of the potential to return to the exhaust
core for additional Fermi reflections. This behavior is shown by the test particle trajectory  in Fig~\ref{Fig:Ti-too-small}c, and it is shown in the supplemental material that the reflection
 is due primarily to the potential and not to mirroring.  However, electrons lose energy
in their reflection from the potential (in the frame of the x-line)
since the potential is moving outward along the magnetic field. The
energy gain from Fermi reflection continues to exceed the loss from
reflection from the potential as long as the expansion velocity
is less than $v_0$, the field line velocity. Thus,
Fig.~\ref{Fig:phi-effect-on-Ti}a, which demonstrates that the expansion velocity and field line velocity converge downstream, establishes how
electron energy gain is limited. The shock bounding the reconnection
exhaust, as discussed by \cite{Liu12}, carries the
potential outward along ${\bf B}$. The electron temperature increases,
increasing the shock velocity, until the shock speed reaches $v_0$ and
electron heating saturates. 

Here we do not present a complete model of the electron heating during
reconnection, which requires a full understanding of the dependence of
the shock velocity on electron and ion temperatures upstream and
downstream. Instead we simply note that in the limit of low upstream
pressure (high upstream mach number) the propagation speed of a simple
parallel propagating slow shock with a jump in the parallel temperature is $2\sqrt{\Delta (T_e+T_i)/m_i}$. Equating this speed to
$v_0=c_{Aup}$, we find $\Delta (T_e+T_i)=0.25 \ m_i\,c_{Aup}^2 = 0.25\,W,$ 
which is within a factor of two of  the simulation and observational findings of $0.15\,W$ \citep{Phan13,Phan14}.
There is significant uncertainty in the 0.25 coefficient, however, due to the simplistic nature
of the shock analysis used to derive it.  Nevertheless, the basic idea that the electron and ion
temperature increments are linked through their control of the
propagation speed of the shock and associated potential is
consistent with the results of Fig.~\ref{Fig:overview}c.

\section{Conclusions:}

We present the results of PIC simulations of reconnection-driven
electron and ion heating that suggest that the partition of energy
gain of the two species is controlled by the large-scale potential
that develops to prevent hot electrons in the reconnection exhaust
from escaping along open magnetic field lines. We first show that the
relative heating of electrons and ions is controlled by the relative
magnitudes of the upstream temperatures of each species -- high
upstream electron temperature yields much higher electron than ion
heating demonstrating that the typical partition of energy seen in
space and the laboratory are not universal. We then carry out a
detailed study of ion heating and show that the potential slows ions
injected into the exhaust to values below the Alv\'enic exhaust flow speed. Ion heating therefore can fall well below the characteristic value $\Delta
T_i = m_i\,v_0^2/3$ predicted by simple Fermi reflection. The scaling
of $\Delta T_i$ in the simulations is consistent with this theory. The
suppression of ion heating becomes very significant for high upstream
electron temperature when the potential becomes very large. The mechanism by which the potential controls electron heating is also discussed. The
potential confines electrons within the exhaust and enables them to
undergo multiple Fermi reflections. The outward propagation of the
spatial variation of the confining potential, which is linked to the slow shock that bounds the
exhaust, ultimately halts electron energy gain when its velocity
reaches the exhaust velocity -- energy gain through Fermi reflection then
balances energy loss through reflection off the outward propagating
potential. Thus, the electron temperature rises until the
shock/potential velocity matches the exhaust velocity. The potential
is therefore the key ingredient that controls both electron and ion
heating and their relative energy gain. 

An intriguing result is that the total plasma heating $( \Delta T_{tot} = \Delta T_e + \Delta T_i)$ in the simulations is constant with \(\Delta T_{tot} \approx .15\ W\), which is consistent with recent magnetospheric observations~\citep{Phan13,Phan14}. Although this is an exciting result, our simulations explore only the small parameter regime of symmetric and anti-parallel
reconnection. Determination of the generality of the $\Delta T_{tot}$ scaling will require
a more systematic scaling study. Regarding the comparison with satellite observations: On the
one hand the fact that the observations are of asymmetric reconnection and the simulations
are symmetric requires some caution during comparison; Clearly, the simulation scaling
study should be extended to asymmetric reconnection. On the other hand, the fact that asymmetric
observations have such good agreement with symmetric simulations implies that the scaling may be a general
result, applicable to a wide range of reconnecting systems.

\begin{acknowledgments}
 This research was support by the NASA Space Grant program at the
  University of Delaware; NSF Grants Nos. AGS-1219382 (M.A.S) and
  AGS-1202330 (J. F. D); NASA Grants
  Nos. NNX08A083G--MMS IDS (T.D.P and M.A.S), NNX14AC78G (J.F.D), NNX13AD72G (M.A.S.), and
  NNX15AW58G (M.A.S). Simulations and
  analysis were performed at the National Center for Atmospheric
  Research Computational and Information System Laboratory (NCAR-CISL)
  and at the National Energy Research Scientific Computing Center
  (NERSC).  We wish to acknowledge support from the International
  Space Science Institute in Bern, Switzerland.
\end{acknowledgments}

\bibliographystyle{agufull08}

\end{article}

\end{document}